# The parallel logic gates in synthetic gene networks induced by non-Gausssian noise

YONG XU[1], XIAOQIN JIN, HUIQING ZHANG

Department of Applied Mathematics

Northwestern Polytechnical University, Xi'an, 710072, China

**Abstract -** The newly appeared idea of Logical Stochastic Resonance (LSR) is verified in synthetic gene networks induced by non-Gaussian noise. We realize the switching between two kinds of logic gates under optimal moderate noise intensity by varying two different tunable parameters in a single gene network. Furthermore, in order to obtain more logic operations, thus providing additional information processing capacity, we obtain in a two-dimensional toggle switch model two complementary logic gates and realize the transformation between two logic gates via the methods of changing different parameters. These simulated results contribute to improve the computational power and functionality of the networks.

## 1. Introduction

Recently, based on the theory of stochastic resonance (SR) and the cooperation among nonlinear dynamics, noise and weakly driven signals, the logical stochastic resonance (LSR) is firstly proposed by Murali et al. [1], who shows a flexible and reliable logic gate in a bistable system with two adjustable thresholds. As a new idea in nonlinear dynamics, LSR has attracted immense attentions of experts in various fields. And the fundamental OR, AND, NOR, NAND and XOR logic operations are used to the basis of universal computation, so experts tried their best to enhance the reliability of logic gates by exploiting the interplay between noise and nonlinearity. Based on the practical usefulness of logic operations, the logic gates have been widely used in various fields. For instance, in an electronic circuit system, LSR was demonstrated and the logic gates can be allowed to transform from AND into OR by simply varying two thresholds [2]. And Bulsara et al. proved in a bistable optical system that two parallel logic gates can be obtained by adjusting the bias [3]. Besides, the realization of switching between two logic gates can also be demonstrated in a synthetic biological logic block [4], an engineered gene network [5] and synthetic gene networks [6]. However, to our knowledge, previous studies on the switching between two logic gates are assumed that the noise is Gaussian white case. Motivated by these articles, we wonder to know whether the performance of logic gates can be observed in synthetic gene networks driven by a noise source taken as non-Gaussian noise, which has never been reported. Additionally, the non-Gaussian noise is superior to the Gaussian white noise in realizing the switching between two logic gates. That is: the transformation between logic gates can be realized only by varying the system parameter in a nonlinear system induced by Gaussian white noise. However, since the non-Gaussian noise has a particular distribution controlled by the departure parameter, different distribution has strong effect on the logical stochastic resonance phenomenon, so the transformation between logic gates can be obtained both by varying the system parameters and by varying the departure parameter of the non-Gaussian noise.

---

[1] E-mail:hsux3@nwpu.edu.cn. Tel./fax numbers:86-29-88431637.

In this paper, we focus our attention on the switching between two logic gates in a single gene network model and a two-dimensional toggle switch model. And we indicate in the single gene network model that the gate reconfiguration can be obtained by varying a tunable system parameter under an optimal wide range of noise intensity. Moreover, with fixed system parameters, different parameter $p$ values, which denote the degree of the departure from non-Gaussian noise $\eta(t)$ and Gaussian noise, can flexibly lead to two different logic gates. Therefore, the non-Gaussian noise serves as a logic gate selector in this network. Further, we can obtain parallel complementary logic gates simultaneously in a two-dimensional toggle switch model and can also switch one kind of logic gates into another by changing system parameters or noise parameter $p$ under optimal noise intensity. So the non-Gaussian noise plays a crucial rule in gate reconfigurations and these implementations of gates contribute to increase the computational power and functionality of the networks, thus providing a better understanding of the information processing capacity. Additionally, with successfully performing the logic functions (e.g., AND/OR) in synthetic gene networks, we can also use LSR in many different environments and change its behavior.

In this paper, firstly we briefly introduce the statistical properties and the generation of non-Gaussian process $\eta$. Secondly we describe a single gene network model, show numerical results that demonstrate this model can operate a switching between two logic gates through the method of changing the system parameter or the departure parameter $p$, and discuss the reliability of the logic gates. Thirdly we present a two-dimensional toggle switch model, and indicate that two parallel logic gates can be obtained simultaneously. By using the similar methods to the single gene network, we realize the switching between two types of logic gates.

## 2. The generation of non-Gaussian process $\eta$

We consider that the stochastic process $\eta(t)$ is characterized by the following Langevin equation [7]

$$\dot{\eta} = -\frac{1}{\tau}\frac{d}{d\eta}V_p(\eta) + \frac{\sqrt{2D}}{\tau}\xi(t), \qquad (1)$$

where $\xi(t)$ is a standard Gaussian white noise of zero mean and correlation $\langle \xi(t)\xi(t')\rangle = \delta(t-t')$, $V_p(\eta)$ is given by

$$V_p(\eta) = \frac{D}{\tau(p-1)}\ln\left[1+\frac{\tau}{D}(p-1)\frac{\eta^2}{2}\right], \qquad (2)$$

and the statistical properties of the non-Gaussian noise $\eta(t)$ are defined as

$$\langle \eta(t)\rangle = 0, \qquad (3)$$

$$\langle \eta^2(t)\rangle = \begin{cases} \dfrac{2D}{\tau(5-3p)}, & p\in\left(-\infty,\dfrac{5}{3}\right), \\ \infty, & p\in\left[\dfrac{5}{3},3\right), \end{cases} \qquad (4)$$

in which $\tau$ denotes the correlation time of the non-Gaussian noise $\eta(t)$, $D$ denotes the noise intensity of the Gaussian white noise $\xi(t)$. The parameter $p$ is used to control the degree of the departure from the non-Gaussian noise and Gaussian noise. The non-Gaussian noise has a particular distribution: while $p \approx 1$ corresponding to Gaussian noise, for $p>1$ corresponding to a long tail distribution and $p<1$ produces a cut-off distribution.

Next we generate the non-Gaussian noise $\eta(t)$ as follows [8]: first we discrete time in steps of size $h=0.01$, then the value of the non-Gaussian process $\eta(t)$ at the discretized times can be obtained by

$$K(t) = hV_p(\eta(t)), \qquad (5\text{-}1)$$

$$L(t) = \frac{\sqrt{2D}}{\tau} h^{1/2} G(t), \qquad (5\text{-}2)$$

$$\eta(t+h) = \eta(t) + \frac{h}{2}\left[V_p(\eta(t)) + V_p(\eta(t) + K(t) + L(t))\right] + L(t), \qquad (5\text{-}3)$$

with the initial value $\eta(0)=0$. We set $\eta_1 = \sqrt{\frac{2D}{\tau(1-p)}}$, and the values of $G(t)$ for different times are Gaussian random numbers with zero mean and the variance one. In the case $p<1$, it is worth noting that if the generated $G(t)$ value substituted into the Eq. (5) leads $\eta(t+h)$ to go outside the interval $(-\eta_1, \eta_1)$, then we discard the $G(t)$ value and generate a new $G(t)$ until meeting the condition $|\eta(t+h)| < \eta_1$. Whereas for $p \geq 1$, this problem never occurs. In this discretization method the process $G(t)$ is obtained.

**3. The switching between two logic gates in a single gene network model**

To our knowledge, due to the fact that cells are intrinsically noisy biochemical reactors and significant statistical fluctuations in molecule numbers, and reaction rates can be resulted in by low reactant numbers [9], so it is worth studying the role of noise in nonlinear gene system, in particular the noise's relevance to biological computation. For demonstrating the biological LSR and investigating the switching between logic gates, here we explicitly present a single gene network model induced by non-Gaussian noise and use the quantitative model mentioned in [10-11], in which the regulation of the operator region of $\lambda$ phage is described and three operator sites are consisted in the promoter region. This network is given by [6]:

$$\dot{x} = \frac{m(1 + x^2 + \alpha\sigma_1 x^4)}{1 + x^2 + \sigma_1 x^4 + \sigma_1\sigma_2 x^6} - \gamma x + I(t) + D\eta(t), \qquad (6)$$

where $x$ and the integer $m$ denote the concentration of the repressor and the number of plasmids per cell, respectively, the parameter $\gamma$ is considered to be a tunable parameter which is directly proportional to the protein degradation rate, here we take $\sigma_1 = 1.95$, $\sigma_2 = 0.083$, $\alpha = 10.9$ for the operator region of the $\lambda$ phage. Since designing a plasmid with a given copy number is possible, so we take $m = 1$. $\eta(t)$ is a non-Gaussian noise, which has been described in the above section. $I(t) = I_1(t) + I_2(t)$ is a low amplitude input signal, $I_1$ and $I_2$ are two aperiodic trains of square pulses encoding two logic inputs. With no loss of generality, for logic 0, we set $I_{1,2} = 0$, whereas for logic 1 we set $I_{1,2} = 0.75$. Then $I$ takes values 0, 0.75, 1.5, corresponding to the input set $(0,0)$, $(0,1)/(1,0)$ and $(1,1)$. With $(0,1)$ and $(1,0)$ lead to the same $I$ value, thus one can speak that $I$ is a three level aperiodic wave form.

The logic output is determined by the system state, here we interpret the state in the left well as logic 0 and the state in the right well as logic 1. The complementary gate can be defined similarly by interpreting the state in the left well as logic 1 and the state in the right well as logic 0.

The logic gate can be realized in a system when the logic output determined by the system state matches the logic output in the logic truth table (see Table 1). Besides, according to the logic input and logic output stated above, we can ascertain the kind of logic gates.

| Input set $(I_1, I_2)$ | OR | NOR | AND | NAND |
|---|---|---|---|---|
| $(0,0)$ | 0 | 1 | 0 | 1 |
| $(1,0)/(0,1)$ | 1 | 0 | 0 | 1 |
| $(1,1)$ | 1 | 0 | 1 | 0 |

**Table 1** Relationship between two logic inputs and the logic output for the fundamental AND, NAND, OR, and NOR logic behaviors.

The response of the system with fixed $p=0.5$, correlation time $\tau=0.008$, optimal noise intensity $D=0.05$ for $\gamma=5.5$ and $\gamma=6.0$ are shown in Fig. 1 (a) and (b), respectively.

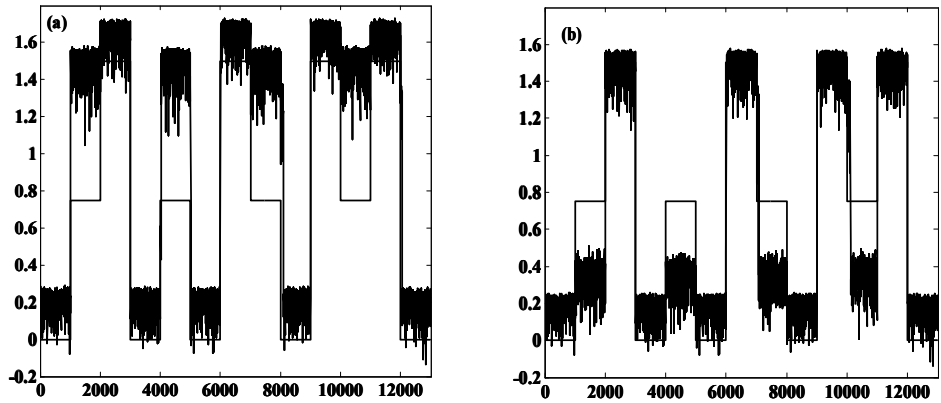

Fig. 1: (a) Response of the system (6) with $\gamma=5.5$. (b) Response of the system (6) with $\gamma=6.0$. The solid line denotes the output stream and the dashed line denotes the input stream. Other parameters in system (6) are: $m=1$, $\sigma_1=1.95$, $\sigma_2=0.083$, $\alpha=10.9$, $p=0.5$, $\tau=0.008$, $I=0, 0.75, 1.5$, optimal noise intensity $D=0.05$. Then a reliable OR/NOR is obtained for $\gamma = 5.5$ and AND/NAND gate is obtained for $\gamma = 6.0$.

Obviously, as can be observed from Fig. 1 that two kinds of logic gates are obtained for two different $\gamma$ values. This can be easily understood from the fact that the parameter $\gamma$ can be used to control the degree of asymmetry in the potential well of the nonlinear system (6). Here we consider the logic output to be logic 0 when the system state $x(t) < x_m$ and logic 1 corresponding to $x(t) > x_m$, $x_m$ is a middle point between the left well and the right well. Then in system (6) with $\gamma = 5.5$, a clear logic gate OR can be obtained. Whereas, in the system with $\gamma = 6.0$, a clear logic gate AND can be realized.

However, we wonder to know how the noise intensity affects the behavior of the system response, and show how the optimal range of noise intensity work in the implementation of the correct logic operation, then we need to define a quantify that help us check the reliability of obtaining a correct logic output. So as a measure of the reliability in the logic gate, the success probability of obtaining expected logic gate is introduced. The probability is calculated as the ratio of the number of correct logic outputs to the total number of runs [12]. A run is considered to be success like that: each run samples over the four input sets $(0,0)$, $(0,1)$, $(1,0)$, $(1,1)$ in different permutations, the logic output obtained from the system response matches the logic output in the truth table for all the four input sets. If this is not the case, we regard the run as a failure. It is apparent that the probability should change as the noise intensity changes. Yielding the optimal performance or obtaining the reliable logic gate is defined when the success probability is close to 1. For the aim of ascertaining the optimal ranges of noise intensity to the correct logic operation. The success probability of obtaining the correct logic gate as a function of the noise intensity $D$ for $\gamma = 5.5$ and $\gamma = 6.0$ are displayed in Fig. 2 (a) and (b), respectively.

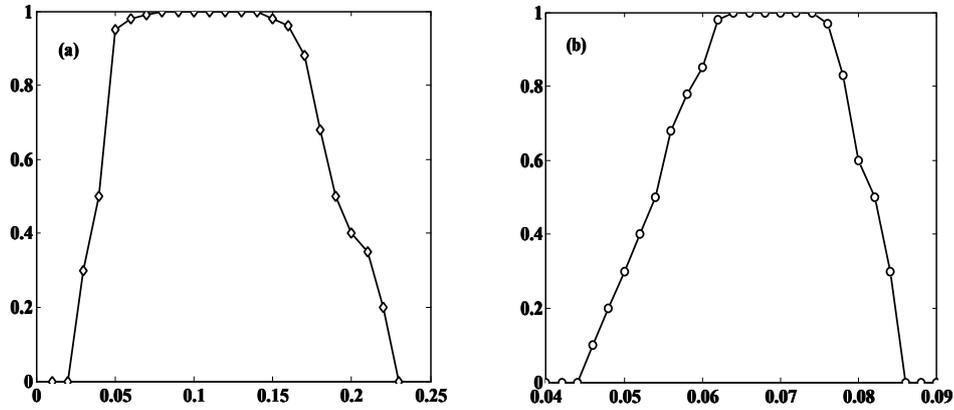

Fig. 2: (a) Success probability $P(OR)$ as a function of the noise intensity $D$ with fixed $\gamma = 5.5$. (b) Success probability $P(AND)$ as a function of the noise intensity $D$ with fixed $\gamma = 6.0$. Other parameters are the same with Fig. 1.

As can be observed in Fig. 2 that the success probability increases with the noise intensity increases, and then reaches the maximum and then decreases, so the fundamental logic gates OR and AND are realized consistently in an optimal window of moderate noise, and now the success probability is close to 1. The remarkable point here is that the correct logic gates are only realized in presence of the noise. What is noteworthy is that there occurs a flat maximum rather than the usual peak associated in stochastic resonance (SR) [13-14]. In a particular case, the correct biologic logic gate cannot be obtained for too small or too large noise intensity, this is because that for too small noise intensity, the external random force acting on the single gene network system is not strong enough to lead the system to change its own stable state. However, for too large noise intensity, the external random force is too strong so that the noise continuously makes the system skip between two wells, thus losing the stability of implementing the correct gate. Besides, the greatest control over the optimal performance in this configuration is the depths of two potential wells, since $\gamma$ can be used to control the degree of asymmetry of the potential, and different values of $\gamma$ lead the system states to fall into different wells, therefore, the logic gate OR can be switched into the logic gate AND by means of varying the system parameter $\gamma$.

We have started out our analysis taking into account the parameter $\gamma$ that leads to the switching between two logic operations. Furthermore, we can also realize the switching between two different types of logic gates by changing the noise parameter $p$. With fixed $\gamma = 6.8$, Fig. 3 demonstrates that the system can realize the logic gate AND when $p = 0.6$ and logic gate OR when $p = 1.3$. The toggle between two logic operations are realized via the varying noise parameter $p$. So random noise plays a crucial role in the implementation of the LSR and it can serve as a logic gate selector.

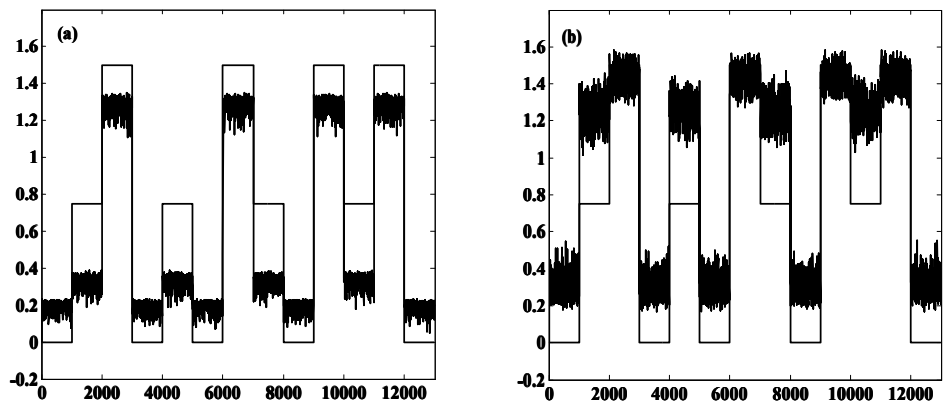

Fig. 3: (a) Response of the system (6) with $p = 0.6$. (b) Response of the system (6) with $p = 1.3$. The dashed line shows the input signals and the solid line shows the desired logic outputs. The solid line denotes the output stream and the dashed line denotes the inputstream. Other parameter values are $m=1$, $\sigma_1 = 1.95$, $\sigma_2 = 0.083$, $\alpha = 10.9$, $\gamma = 6.8$, $\tau = 0.008$, $I = 0, 0.75, 1.5$, $D = 0.05$. It is clear that the logical AND is obtained in (a) and OR is obtained in (b).

In order to indicate how two different types of logic gates depend on the noise parameter $p$ and check the robustness of the LSR, we show the performance versus the noise intensity $D$ and the noise parameter $p$ for the logic gate AND and OR in Fig. 4.

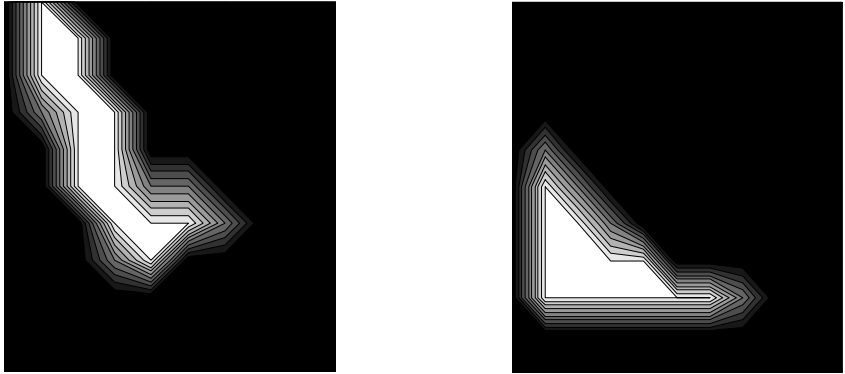

Fig. 4: (Color online). (a) Performance of the logic gates AND vs. noise intensity $D$ and $p$. (b) Performance of the logic gates OR vs. noise intensity $D$ and $p$, other parameters as in the text. (The light portion denotes the logic gate can be obtained reliably).

From Fig. 4 we can observe that there are different ranges of noise parameter $p$ to obtain different logic gates. And it is evident that the fundamental logic gates AND and OR are obtained consistently in an optimal window of noise intensity. As for this phenomenon, the reason is that different noise parameter $p$ can also cause the system states to fall into different wells, so we can make a conclusion that the noise parameter $p$ can also lead to the switching between two logic gates. Besides, it is worth noting in this single gene network that the in the case $p \approx 1$, any logic gate cannot be realized. But for both $p<1$ and $p>1$, the logic gate can be realized. Specifically, the logic gate AND can be obtained with $p<1$ and the logic gate OR can be obtained with $p>1$. As discussed for the logical stochastic resonance phenomenon induced by the non-Gaussian noise, we can find that a strong effect in the realizing of logic gate has produced due to the departure from the Gaussian distribution.

**4. The switching between two logic gates in a two-dimensional toggle switch model**

In order to highlight the possibility of processing two logic operations in parallel in a high-dimensional gene network, we consider a two-dimensional toggle switch model, which is composed of two repressors and two constructive promoters [15]. The opposing promoter can transcribe the repressor that is used to inhibit each promoter. We can describe this model in the following form [16]:

$$\dot{X} = \frac{\alpha_1}{1+Y^{n_1}} - d_1 X + g_1, \qquad (7a)$$

$$\dot{Y} = \frac{\alpha_2}{1+X^{n_2}} - d_2 Y + g_2, \qquad (7b)$$

in which $X$ and $Y$ are the concentration of repressor R1 and R2, respectively. The effective rates of synthesis of repressor R1 and R2 are denoted as $\alpha_1$ and $\alpha_2$. And the cooperativity of repression of promoter 2 (promoter 1) is $n_1(n_2)$. The degradation rates and the basal synthesis rates are denoted as $d_i(i=1,2)$ and $g_i(i=1,2)$, respectively. When fixed the system parameters $\alpha_1=\alpha_2=5$, $n_1=n_2=1.9$, $d_1=0.9$, $d_2=1$, $g_1=0.045$, $g_2=0.1$, then the geometric structure of the deterministic Eq. (7)

illustrated in Fig. 5 indicates the origin of the bistability [15]. As can be seen from Fig. 5 that the nullclines of $dX/dt = 0$ and $dY/dt = 0$ intersect at three points due to their sigmoidal shape, giving rise to two stable steady states and one unstable state. The structure of the toggle switch network produces two parts: one part above the separatrix is the stable region of state 1; the other part below the separatrix is the stable region of state 2.

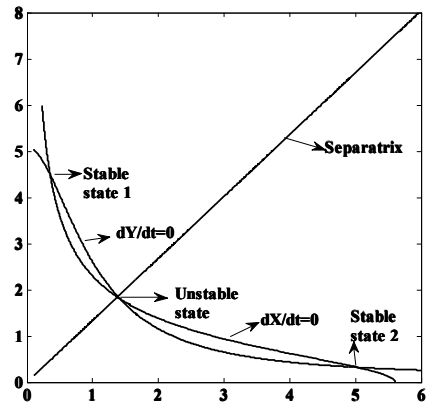

Fig. 5: Null-clines for the determinstic system (7) in the phase space. The system is a bistable potential function with the parameters: $\alpha_1 = 5$, $\alpha_2 = 5$, $n_1 = n_2 = 1.9$, $d_1 = 0.9$, $d_2 = 1$, $g_1 = 0.045$, $g_2 = 0.1$.

For the aim of investigating how non-Gaussian noise introduced into degration rates affects the genetic toggle switch system, we replace the degration rates $d_1$ and $d_2$ in the models (7) with $d_1 + \eta_1(t)$ and $d_2 + \eta_2(t)$, respectively. As the same with the above single gene network model, $\eta_{1,2}$ are non-Gaussian noise with zero-mean, noise intensity $D$, and correlation time $\tau$. Further, the input signal $I = I_1 + I_2$ is considered to drive the Eq. (7b). In such a manner, we can get the following model:

$$\dot{X} = \frac{\alpha_1}{1+Y^{n_1}} - \left(d_1 + \eta_1(t)\right)X + g_1, \tag{8a}$$

$$\dot{Y} = \frac{\alpha_2}{1+X^{n_2}} - \left(d_2 + \eta_2(t)\right)Y + g_2 + I(t), \tag{8b}$$

where the driving input signal $I(t) = I_1(t) + I_2(t)$ is a three-level square wave signal, without loss of generality, here we take $I_{1,2} = 0$ for logic input 0 and $I_{1,2} = 0.1$ for logic input 1. Here we fix the system parameters as

$$\alpha_1 = 5, \ \alpha_2 = 5, \ d_1 = 0.9, \ d_2 = 1, \ g_1 = 0.045, \ g_2 = 0.1, \tau = 0.008, \tag{9}$$

next we show the switching between two kinds of complementary logic gates by changing the value of $n_{1,2}$ and the noise parameter $p$ or varying the value of $g_1$ and $g_2$.

In Fig. 6 we show the input and output streams of system (8) with respect to two variables $X(t)$ and $Y(t)$ with $n_1 = n_2 = 1.6$, $p = 0.5$, then it is available to obtain the correct complementary logic gates NAND for the variable $X(t)$ and AND for the variable $Y(t)$ with the optimal noise intensity $D = 0.09$. By varying the system parameter $n_1 = n_2 = 1.9$ and the noise parameter $p = 1.5$, the 2D system can realize the parallel logic gates NOR for the system variable $X(t)$ and OR for $Y(t)$ with the optimal noise intensity $D = 0.06$, which can be seen in Fig. 7. The transformation between two kinds of parallel logic operations can be achieved, that means, one can tune the shape of the potential wells through changing the system parameter $n_{1,2}$ and the noise parameter $p$, and then different values of these parameters lead the 2D system (8) to be driven into different wells.

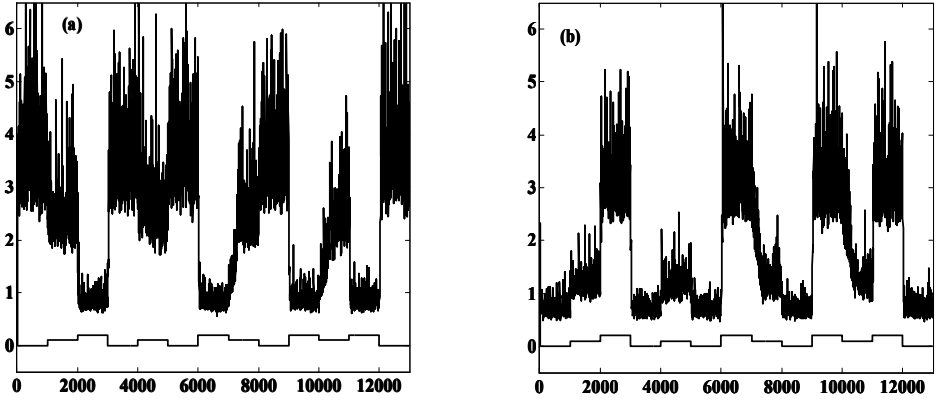

Fig. 6: The input (dashed line) and output (solid line) streams showing two complementary gates with $n_1 = n_2 = 1.6$, $p = 0.5$ and the optimal noise intensity $D = 0.09$. (a) NAND gates with respect to the variable $X(t)$; (b) AND gates with respect to the variable $Y(t)$.

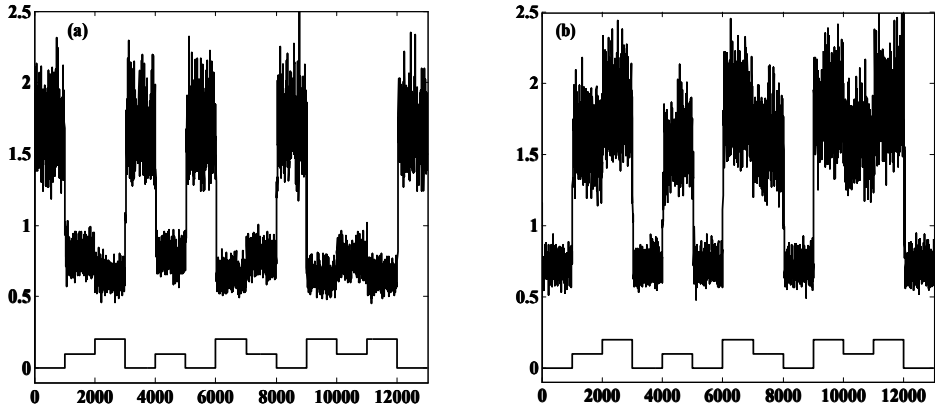

Fig. 7: The input (dashed line) and output (solid line) streams showing two complementary gates with $n_1 = n_2 = 1.9$, $p = 1.5$ and the optimal noise intensity $D = 0.06$. (a) NOR gates with respect to the variable $X(t)$; (b) OR gates with respect to the variable $Y(t)$.

For quantifying the reliability of obtaining the logic gate NAND in Fig. 6 and NOR gate in Fig. 7 with respect to the variable $X(t)$, we use the success probability as a measure and demonstrate in Fig.8 and Fig.9 how the success probability $P(NAND)$ and $P(NOR)$ change with the change of noise intensity. It is clearly evident that in a wide range of moderate noise, the system yields the logic operations NAND and NOR with near certain reliability, i.e., $P(NAND) \sim 1$ or $P(NOR) \sim 1$. Therefore the logic gates perform well over a wide window of moderate noise.

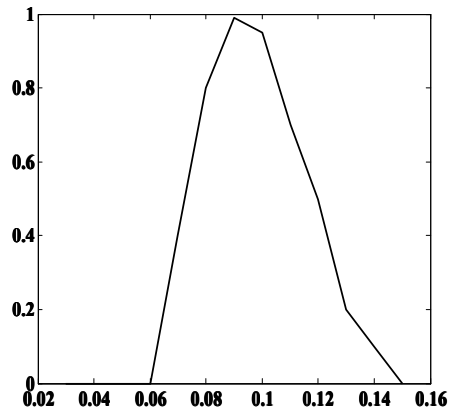

Fig. 8: The success probability of obtaining the logic gate NAND vs. noise intensity $D$. With the system parameters: $n_1 = n_2 = 1.6$, $p = 0.5$, $\alpha_1 = 5$, $\alpha_2 = 5$, $d_1 = 0.9$, $d_2 = 1$, $g_1 = 0.045$, $g_2 = 0.1$, $\tau = 0.008$.

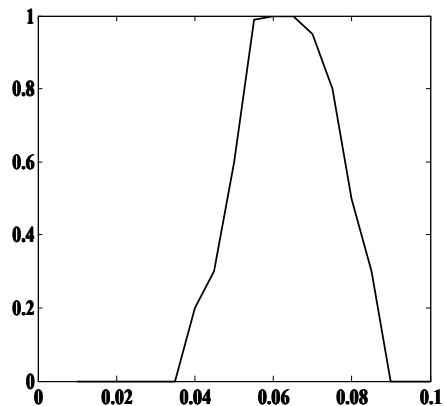

Fig. 9: The success probability of obtaining the logic gate NOR vs. noise intensity $D$. With the system parameters: $n_1 = n_2 = 1.9$, $p = 1.5$, $\alpha_1 = 5$,

$$\alpha_2 = 5,\ d_1 = 0.9,\ d_2 = 1,\ g_1 = 0.045,\ g_2 = 0.1, \tau = 0.008.$$

Additionally, we also can use another method to control the transformation between two kinds of complementary logic gates by varying the value of $g_1$ in the system (8). Leaving all the other system parameters unchanged and same as in Fig. 6 but only change the value of $g_1$ by $g_1 = 0$, then one can realize the logic gate NOR for the variable $X(t)$ and OR for $Y(t)$ in Fig. 10. Namely, the switching between the logic gate NAND (AND) and NOR (OR) can be determined by simply varying the $g_1$ value, which changes the well shape.

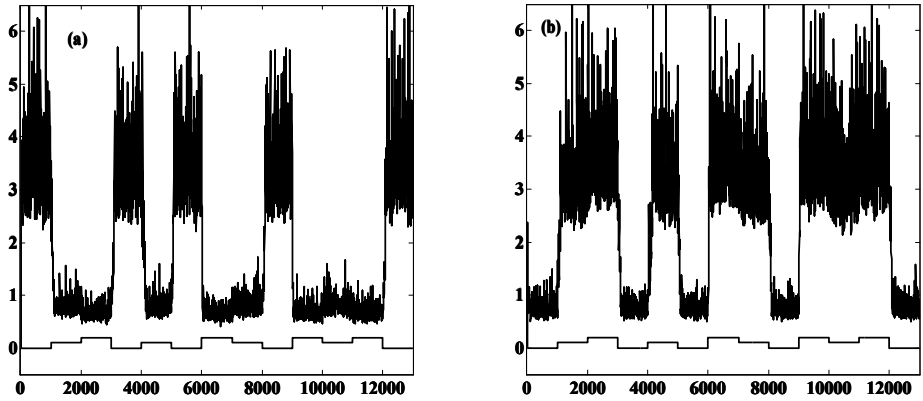

Fig. 10: The input (dashed line) and output (solid line) streams showing two complementary gates with $g_1 = 0$ and the optimal noise intensity $D = 0.09$. (a) NOR gates with respect to the variable $X(t)$; (b) OR gates with respect to the variable $Y(t)$.

Further, in the case $g_1 = 0$ and the logic gate NOR obtained with respect to the variable $X(t)$, we can also quantify its reliability by calculating the success probability. The correct probability $P(NOR)$ as a function of noise intensity $D$ is plotted in Fig. 11. As can be observed from Fig. 11, the system (8) only is able to yield clear logic gate NOR with $P(NOR) \sim 1$ at a certain window of moderate noise strength.

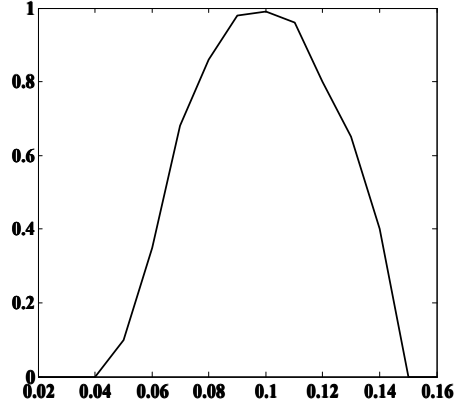

Fig. 11: The success probability of obtaining the logic gate NOR vs. noise intensity $D$. With the system parameters: $n_1 = n_2 = 1.6$, $p = 0.5$, $\alpha_1 = 5$, $\alpha_2 = 5$, $d_1 = 0.9$, $d_2 = 1$, $g_1 = 0.0$, $g_2 = 0.1, \tau = 0.008$.

Above, the three figures (Figs. 6, 7, 10) demonstrate a fact that our biological reconfigurable logic gate performs well under optimal noise intensity. Using two methods to implement a logic block that switches from NAND to the NOR gates with respect to one variable $X(t)$, and the complementary logical output (from AND to OR gates) with respect to the dynamics of other variable $Y(t)$ can be yielded simultaneously.

5. Conclusions

In this paper, we were successfully in varying the phenomenon of logical stochastic resonance in a single gene network model and a two-dimensional toggle switch model induced by non-Gaussian noise. We have demonstrated in a single gene network model that the switching from logic gate AND to OR can be performed by using two methods: one method is changing the system parameter $\gamma$ and the other one is by tuning the departure parameter $p$. Furthermore, we have also discussed the reliability of the logic gates and found that well performed logic gates can be realized in an optimal wide range of noise intensity. In particular, in order to obtain more logic gates and improve the processing capability of computation, we have extended our research to a two dimensional model. We have indicated in a two-dimensional toggle switch model that two parallel complementary logic gates (NAND-AND or NOR-OR) can be obtained simultaneously, and the switching from NAND/AND to NOR/OR gates can be realized through two ways: using the system parameter $n_{1,2}$ and noise parameter $p$ as a logic response controller, or using the system parameter $g_1$ as a logic response controller.

In summary, our results in the synthetic networks have demonstrated the generic nature of LSR. The implementation of parallel logic gates shown in the networks have provided the information processing capability and have the potential to design the biological logic block.

Acknowledge

This work was supported by the NSF of China (Grant Nos. 10972181, 11102157), Program for NCET, the Shaanxi Project for Young New Star in Science & Technology, NPU Foundation for Fundamental Research and SRF for ROCS, SEM.